\documentclass[3p]{elsarticle}
\usepackage{lineno}
\usepackage{placeins}
\modulolinenumbers[5]
\journal{-}
\date{}
\begin{document}
\begin{frontmatter}

\title{Applying physics-based loss functions to neural networks for improved generalizability in mechanics problems}
\author[BioE]{Samuel J. Raymond\corref{mycorrespondingauthor}}
\cortext[mycorrespondingauthor]{Corresponding author}
\ead{sjray@stanford.edu}
\author[BioE,Neuro,MechE]{David B. Camarillo}
\address[BioE]{Department of Bioengineering, Stanford University, Stanford, CA, 94305, USA}
\address[Neuro]{Department of Neurosurgery, Stanford University, Stanford, CA, 94305, USA.}
\address[MechE]{ Department of Mechanical Engineering, Stanford University, Stanford, CA, 94305,USA}
\begin{abstract}
Physics-Informed Machine Learning (PIML) has gained momentum in the last 5 years with scientists and researchers aiming to utilize the benefits afforded by advances in machine learning, particularly in deep learning. With large scientific data sets with rich spatio-temporal data and high-performance computing providing large amounts of data to be inferred and interpreted, the task of PIML is to ensure that these predictions, categorizations, and inferences are enforced by, and conform to the limits imposed by physical laws. In this work a new approach to utilizing PIML is discussed that deals with the use of physics-based loss functions. While typical usage of physical equations in the loss function requires complex layers of derivatives and other functions to ensure that the known governing equation is satisfied, here we show that a similar level of enforcement can be found by implementing more simpler loss functions on specific kinds of output data. The generalizability that this approach affords is shown using examples of simple mechanical models that can be thought of as sufficiently simplified surrogate models for a wide class of problems.
\end{abstract}

\end{frontmatter}


\section{Introduction}
Physics-Informed Machine Learning (PIML) is a cutting-edge new field that sits at the intersection of scientific computing and machine learning. The field is only a few years old now but has already begun producing some valuable insights into the combined approaches of these two domains, particularly in the intersection of computational mechanics, modeling real-world materials/fields, and deep learning, using advanced neural network architectures. 

During the immense rise to power of Artificial Intelligence and Machine Learning in the last two decades, scientific computing was not a largely looked at field of application. Web traffic\cite{Bu20092}, customer habits \cite{Vazquez201468}, Geo-spatial information \cite{Cheng201611}, medical imaging \cite{Litjens201760}, and many others were far easier to apply techniques such as deep learning to, as the guiding models for these areas are often too complex to develop from first principles, and the amount of data available made the training and testing very attractive. Also, as a fundamentally probabilistic endeavor, deep learning may infer values that might violate or exceed reasonable bounds \cite{Su2019828}, which would violate the Laws used in numerical physics, for example, rendering it fundamentally flawed as an application. Scientific data and numerical models, however, are not perfect, and errors and systematic biases are present in these approaches as well. So, if the bounds of expected errors that were to come from a learned-model were similar to those present in a scientific model, an argument could be made to rely just as much on the data-driven approach. In numerical modeling, the starting point is almost always the discretization of the governing partial differential equation (PDE), into a form that can be converted into the syntax of a computer program. On its own, the PDE is in some sense useless as a tool for modeling, until specific boundary and initial values are used, and appropriate parameters set. This transforms the universally applicable PDE to a specific model of a specific use case. In a corollary manner, a deep neural network can be thought of as a universal function approximator \cite{Hornik1991251}, where, with enough connections and neurons, any equation or transfer function relating two variables is theoretically constructable. However, without data and training, the neural network, like the PDE is also completely useless and offers no context-relevant inference. Only once the data has been fed to the neural network and the weights updated through the training process is the general function approximator converted to something far more brittle, yet useful. Brittle here refers to the extent with which the network can make accurate inferences. Only within the bounds of the data used to train the model can a network be relied upon to provide any useful information. Even within these bounds though, there has already been a number of useful applications of deep learning in the physical sciences \cite{raymond2020deep,montgomery2020shale,zhan2020deep,domel2021,raymond2020towards,collins2020acoustic,zhan2020prediction,liu2021time} and this manner of application will continue to develop as more data is generated for scientific purposes. Relying on a data-driven model to provide mechanistic predictions though, creates a number of problems. Adversarial neural network research \cite{Ganin2016} has shown just how brittle these models can be even with human-imperceptible changes to an input. One advantage of numerical models is that they are far less opaque to their brittleness and generally do far better at coping with a large range of input parameter variations without needed to remake the model, though there is certainly a limit to all models. A dangerous prospect, therefore, presents itself when relying on traditional neural networks to provide scientifically relevant inferences. 

The goals of PIML are to address these shortcomings of traditional neural networks, and to leverage the speed and fusion of data that using neural networks affords. Computational models are commonplace in essentially every area of human endeavor, with new methods and techniques constantly introduced to manage increasingly complex scenarios \cite{raymond2014meshfree,raymond2015coupled,raymond2015estimation,raymond2016strategy,raymond2019modeling,wieghold2019detection,	raymond2020fracture}. However, as these models become more sophisticated, they also become horrendously expensive on the scales of climate modeling\cite{Harris2014623} or large-scale molecular dynamics models \cite{Meyers2006427}, and with ever-larger supercomputers requiring even more energy to run\cite{Springer2006230}, PIML also offers a more energy-efficient and sustainable approach to infuse these networks with the benefits of pre-existing domain knowledge. In the last few years, a number of distinct fields of PIML have started to emerge, each using a different approach to embed domain knowledge into deep learning frameworks. These include Physics-Informed Neural Networks (PINNs)\cite{RAISSI2019686}, synthetic data\cite{raymond2020deep}, and data-driven equation learning \cite{Zanna2020}. 

Of these fields this work is primarily interested in the development and use of PINNs \cite{GOSWAMI2020102447,HAGHIGHAT2021113552,HAGHIGHAT2021113741,HALL2021110192,HE2020103610,JIN2021109951,LIU2020113402,RAISSI2019686,WANG2021109914,YANG2021109913,ZHANG2021100220,ZOBEIRY2021104232,doi:10.1122/8.0000138}. The approach in PINNs is to use as an input-output pair to the neural network a fundamental variable, or set of variables for the physical problem, and use these variables to construct the quantities that appear in the governing equation of motion, such as the velocity and pressure fields in a fluids problem. The equation of motion, for instance the Euler or Navier-Stokes equation, is then used as an additional term in the loss function of the deep learning architecture using an additional layer after the output neurons to convert the variables into the forms needed for the equation of motion. The seminal work on this was conducted by Raissi et al. \cite{RAISSI2019686} and the interested reader is encouraged to read their work. The new loss function of PINNs then acts in a similar manner to the residuals in a Finite Element scheme, and the training is performed using traditional backpropagation methods until the loss associated with both the L2-norm and the equation of motion residual is minimized. More work though, is required to understand the right trade-off between these two errors and if one should be weighted more than the other. 

The loss function in PINNs, as in all deep learning, plays a crucial role in the ability of the network to train well \cite{Johnson2016694}. The approach of PINNs is to leverage the full understanding of the governing physics at play to build a basis function, in the form of a neural network, that solves these equations. However, there are many situations where either the governing equation is not clear, or the inclusion of the full PDE introduces such a computational overhead that any efficiencies afforded by the use of a PINN are negated. It may also be that the entire PDE is not required for a sufficiently accurate solution to be produced, for the tolerated error. For example, in viscous flows or simple fluid problems, the full 3D Navier-Stokes is not required and training with such a complex PDE may be unnecessary. In this work we introduce a new approach to this problem by utilizing the Laws of physics in the loss functions, ones that tend to sit above governing PDEs but can be applied much more efficiently and liberally to a number of different problems. To present this, a simpler loss-function approach is presented that uses the conservation of energy as the guiding principle and this is used to build a neural network to predict the motion of a pendulum. This is compared with the traditional, data-only approach, and the discussion of the comparison concludes the work.

\section{Methods}
\subsection{Predicting Motion with a Physics-based loss function}
To compare the efficacy of standard loss functions and PIML-motivated loss functions, the prediction of a simple pendulum's motion is shown for both the conventional and physics-based loss functions. Different starting positions and the resulting predicted motions are show for three different starting angles.
\subsubsection{Conventional Loss Function}
\begin{figure}
	\centering
	\includegraphics[width=0.76\linewidth]{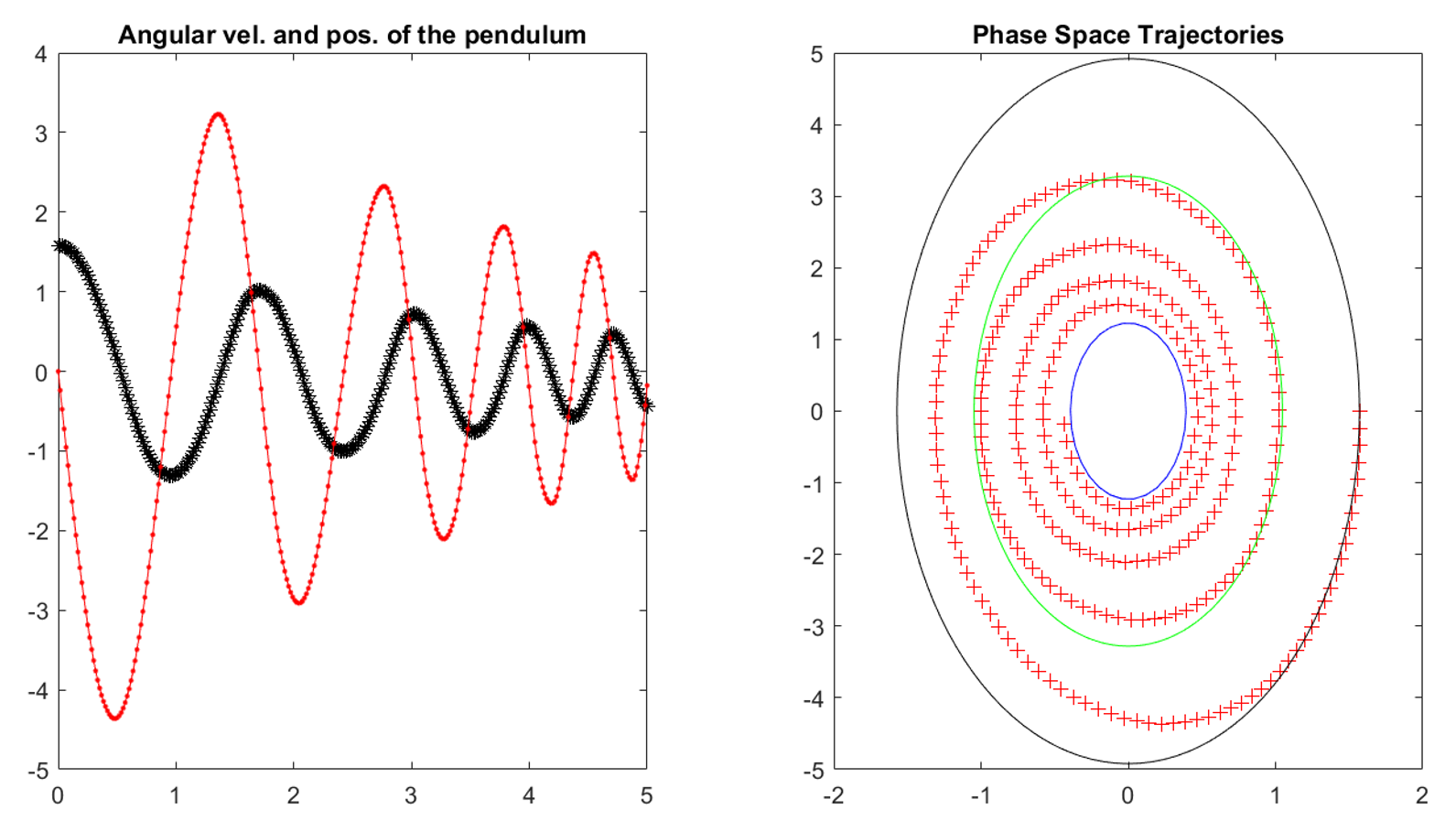}
	\caption{Prediction of the motion of a pendulum from the starting position of $\frac{\pi}{2}$ using the conventional loss function over 4 periods.}
	\label{convpend1}
\end{figure}

\begin{figure}
	\centering
\includegraphics[width=0.76\linewidth]{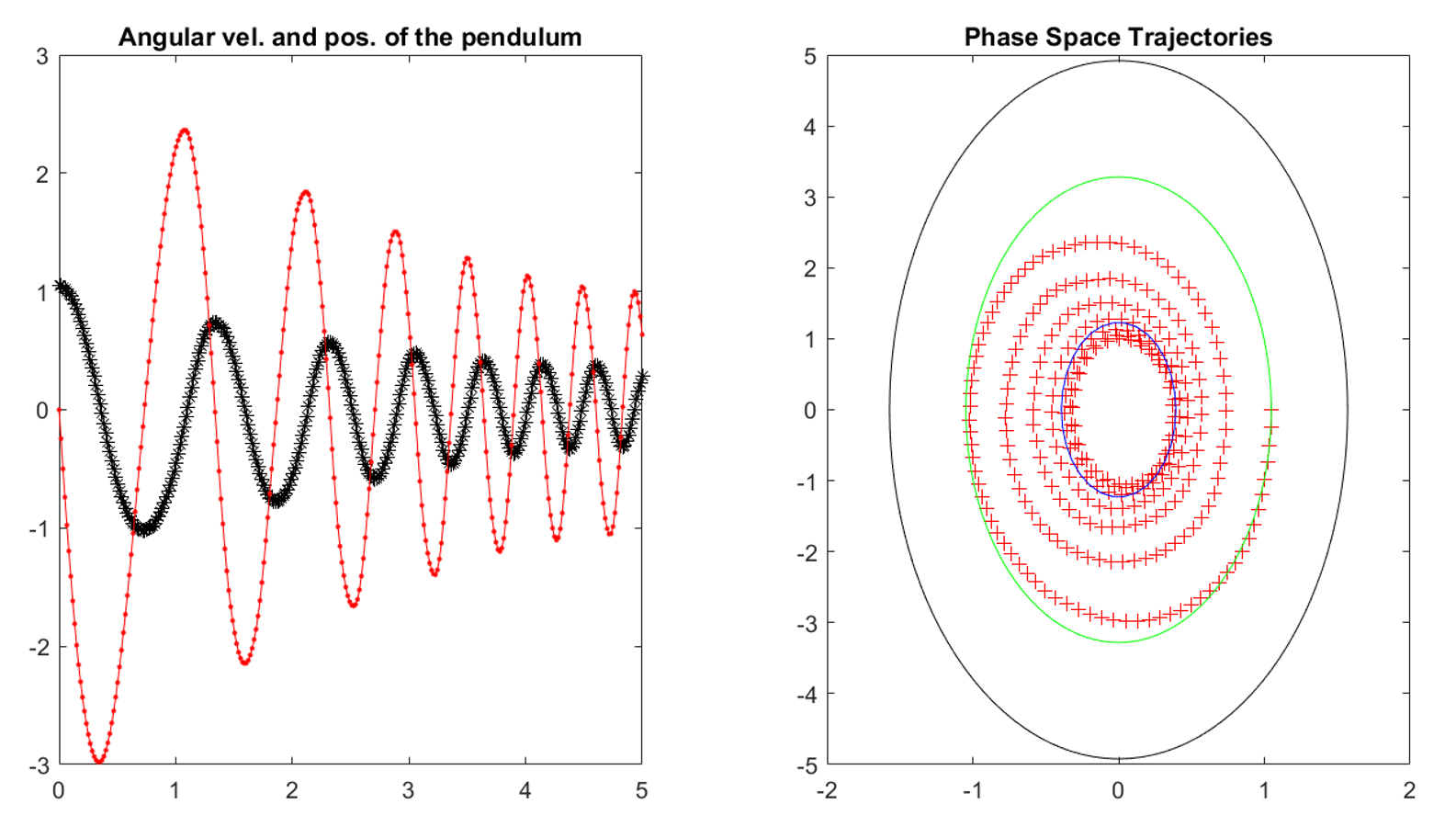}
\caption{Prediction of the motion of a pendulum from the starting position of $\frac{\pi}{3}$ using the conventional loss function over 4 periods.}
\label{convpend2}
\end{figure}

\begin{figure}
	\centering
\includegraphics[width=0.76\linewidth]{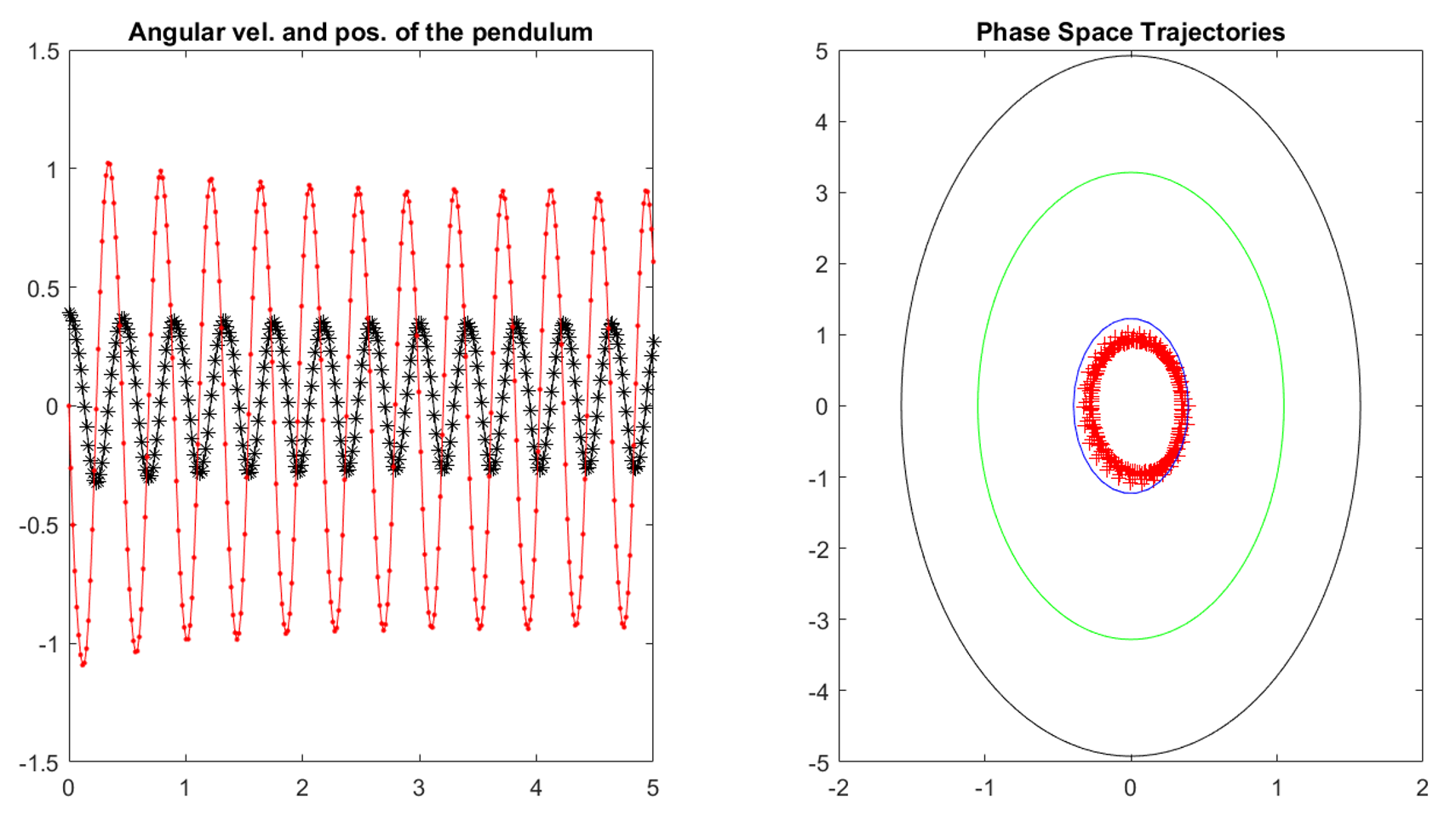}
\caption{Prediction of the motion of a pendulum from the starting position of $\frac{\pi}{6}$ using the conventional loss function over 4 periods.}
\label{convpend3}
\end{figure}

\figurename s \ref{convpend1} - \ref{convpend3} show the results of the prediction of the trajectory of the pendulum from the different starting positions. While the initial predictions follow the phase-space curves for some time, they quickly decay to the smallest energy phase-space.
\subsubsection{Physics-based Loss Function}
\begin{figure}
	\centering
\includegraphics[width=0.76\linewidth]{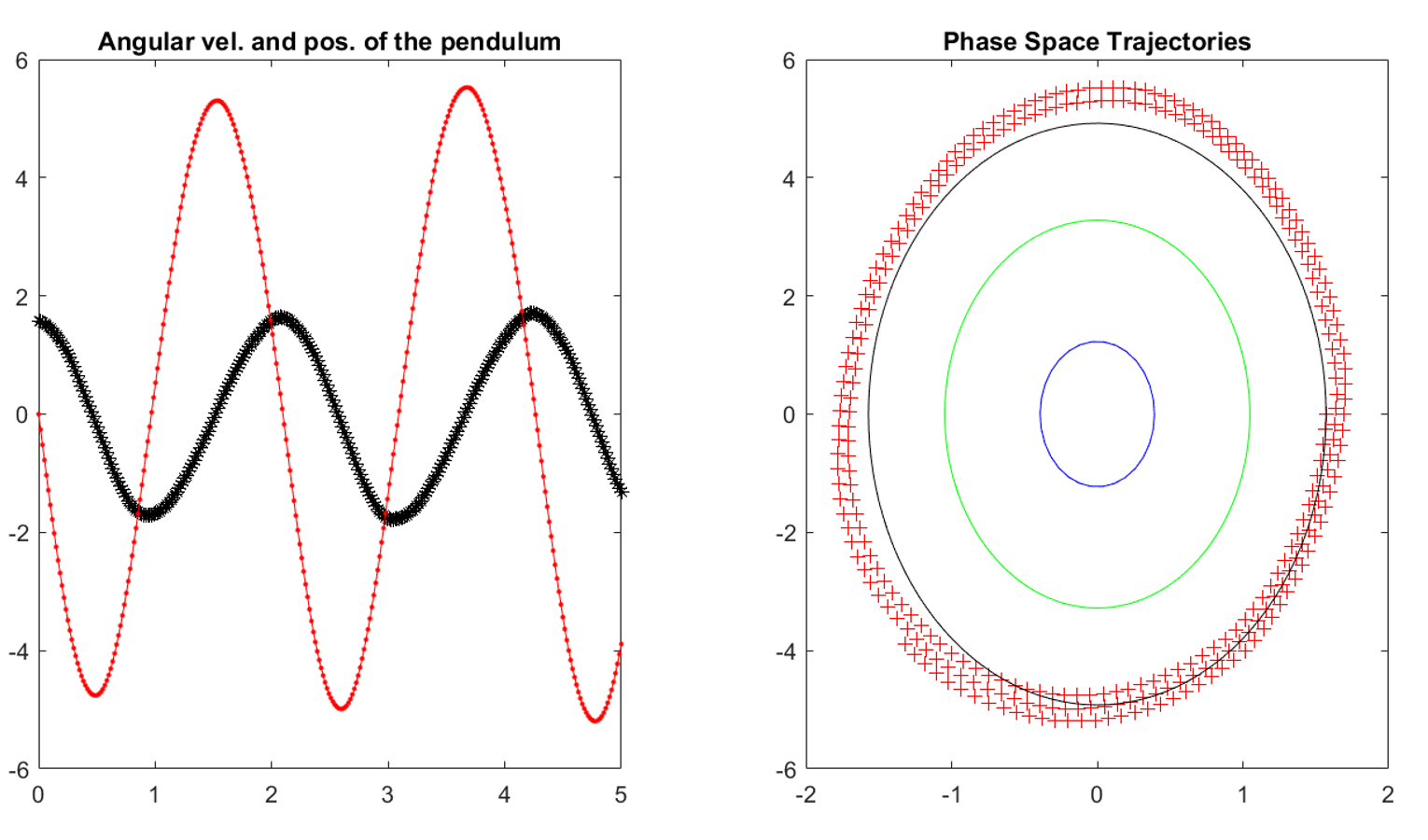}
\caption{Prediction of the motion of a pendulum from the starting position of $\frac{\pi}{2}$ using the combined physics-based and conventional loss function over 4 periods.}
\label{pimlpend1}
\end{figure}

\begin{figure}
	\centering
	\includegraphics[width=0.76\linewidth]{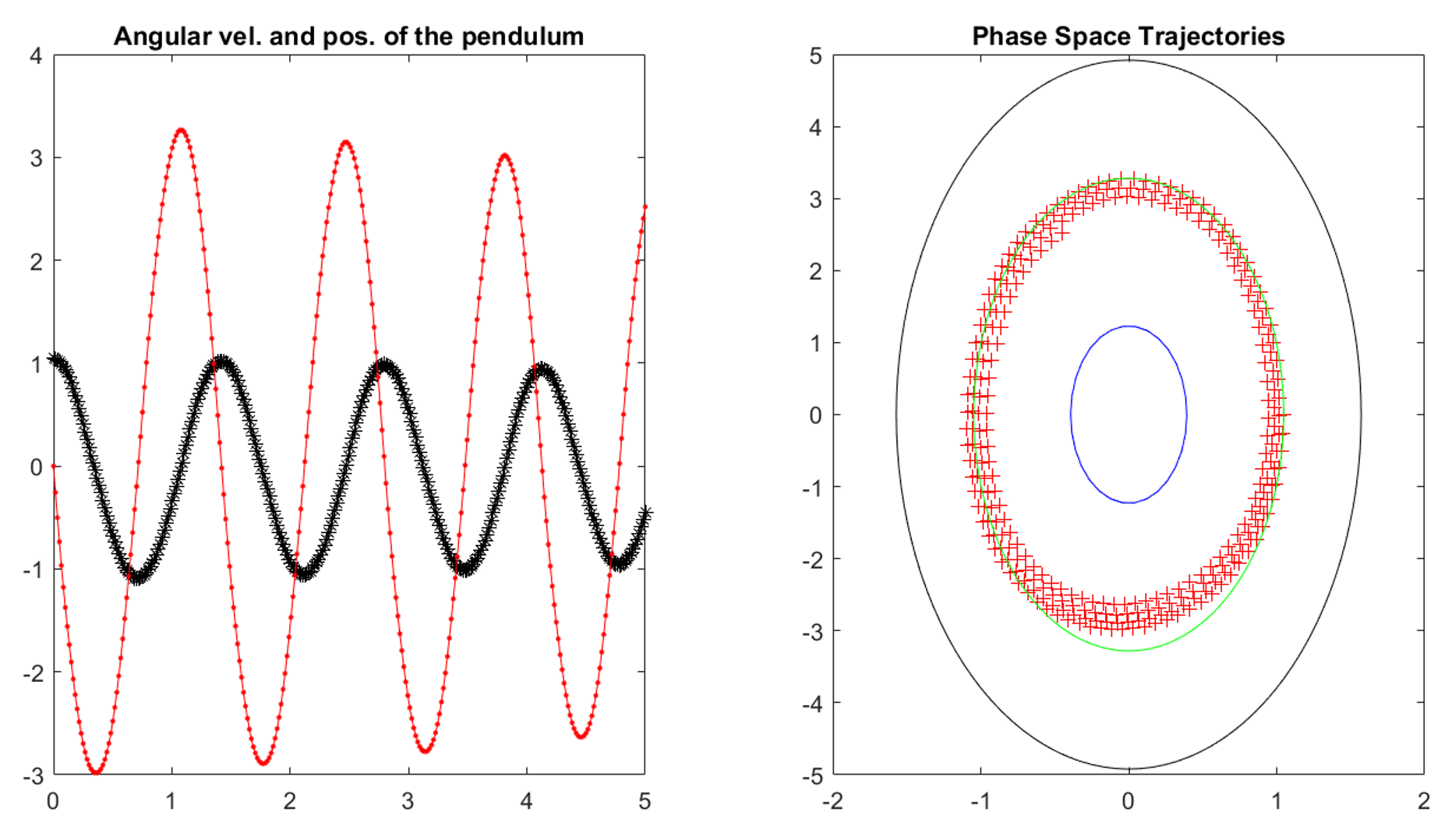}
	\caption{Prediction of the motion of a pendulum from the starting position of $\frac{\pi}{3}$ using the combined physics-based and conventional loss function over 4 periods.}
	\label{pimlpend2}
\end{figure}

\begin{figure}
	\centering
	\includegraphics[width=0.76\linewidth]{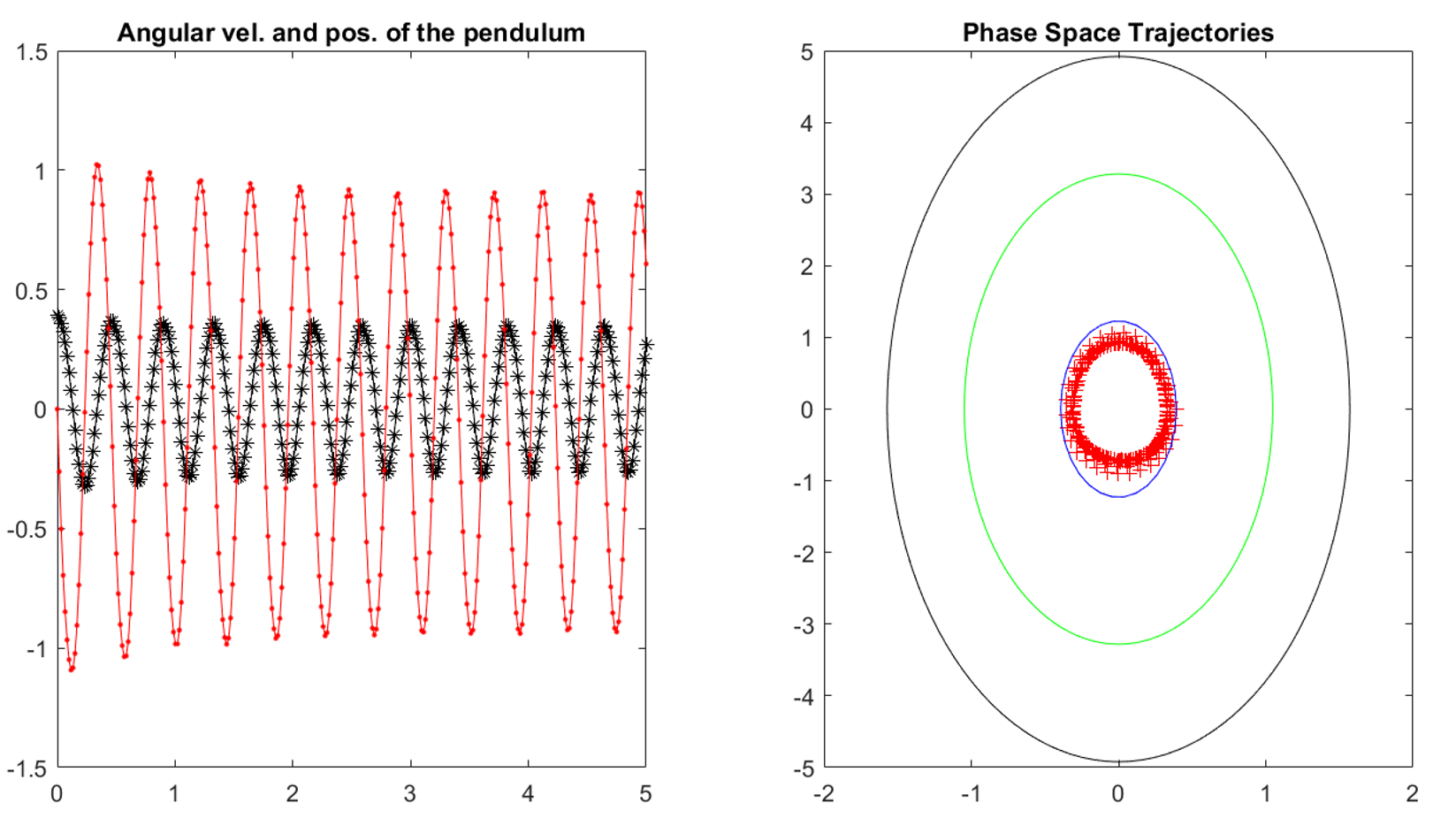}
	\caption{Prediction of the motion of a pendulum from the starting position of $\frac{\pi}{6}$ using the combined physics-based and conventional loss function over 4 periods.}
	\label{pimlpend3}
\end{figure}

\figurename s \ref{pimlpend1} - \ref{pimlpend3} show the results of the prediction of the trajectory of the pendulum from the different starting positions using the neural network trained with the custom loss function adding the conservation of energy term. These results preserve the energy far better than the conventional approach and no decaying of the predictions is found for multiple periods.

\FloatBarrier
\section{Discussion}
In this work a new approach to PIML-based neural networks was proposed that suggests adding only simple but universally applicable laws to the error function to enforce physically accurate predictions.
A neural network was constructed to predict the next time step of motion for a pendulum system, rather than the entire time-space history. This was intended to reflect a more useful use of this form of PIML loss function as only temporally-neighboring states of a system are needed for data in training, rather than the entire time history. A pendulum was modeled and, initially, a regular loss function using the mean-squared error was used to predict the time evolution of a pendulum when raised from a height of $\frac{\pi}{2}$ (\figurename~\ref{convpend1}), $\frac{\pi}{3}$ (\figurename~\ref{convpend2}), and $\frac{\pi}{6}$ (\figurename~\ref{convpend3}). These results showed that the trained network was preferential to the lowest energy data and, after some initial success in prediction of the larger energy systems, decayed to the smallest energy state. This form of mode collapse is likely due to the prevalence in the training to minimize the magnitude of the error and the error itself, preferentiating the lowest valued data. When a new loss function that combined the standard mean-squared error with a measure of the difference in the energy between the input and output states was used, however, the results were drastically different. In each of the three cases (\figurename s \ref{pimlpend1} - \ref{pimlpend3}) the neural network was able to predict the motion of the pendulum and conserve the energy far better than in the original case. It is worth noting that the values predicted by the second neural network in this case show a slightly less precise prediction as the values lie above or below the exact solution, but this is likely the effect of having to balance both this extra energy conservation constraint, and the minimization of the error in predicted values.\\

Despite the simplistic models used to test the proposed architecture, these results show a promising new avenue to explore when wanting to use simulated data and/or real-world data for physically relevant predictions and inferences. Adding new constraints such as the conservation of energy into the loss function is far more computationally efficient than the application of the traditional PINN approach and can be essentially universally applied to any system where the energy is conserved, or the loss to the system can be quantified appropriately. This also implies that other conservation law, linear and angular momenta, and even other forms of symmetries could be embedded to a loss function when the input and output neuron variables are physical quantities. However, as most models of interest do comprise of the same energy forms presented here, systems of multiple particles, and more complex dynamics may pose more difficult to predict as the number of phase-space dimensions increases. In this work only simulated data, which is often not a good measure of what can be expected of real-world data, was used meaning that the networks may not perform as well when managing real-world noisy data. There is also the need for a large amount of training data that necessitates a large computational load to be performed before the training of the network can be done effectively. Also, while creating this data, the choice of the parameters and their ranges is a problem when considering how wide the spectrum of values would need to be for a neural network-based approach to be useful. However, as the networks are much faster to use once they are trained, and with the immense amount of existing simulated and real world data that already exists and is oftentimes just discarded anyway, the cost-benefit analysis looks promising for this form of PIML moving forward.

\section{Conclusion}
In this work we present a new form of physics-based loss function to train a neural network to prediction the evolution of a system. Unlike conventional loss functions that only consider the error between the predicted and true value, here we show that in cases where systems of different energy are used as training data, conventional loss functions fail to properly predict different systems. Instead, we add an additional term to the error to enforce the conservation of energy between the input and the out of the network and the results show that this drastically improves the prediction of the evolution of different systems with differing levels of total energy. This simple but powerful alteration to the loss function in the field of Physics-Informed Machine Learning opens up a new set of opportunities to fuse simulation and real-world data for deep learning predictions of physical systems.

\bibliography{mybibfile}

\begin{thebibliography}{10}
\expandafter\ifx\csname url\endcsname\relax
  \def\url#1{\texttt{#1}}\fi
\expandafter\ifx\csname urlprefix\endcsname\relax\def\urlprefix{URL }\fi
\expandafter\ifx\csname href\endcsname\relax
  \def\href#1#2{#2} \def\path#1{#1}\fi

\bibitem{Bu20092}
X.~Bu, J.~Rao, C.-Z. Xu, A reinforcement learning approach to online web
  systems auto-configuration, 2009, pp. 2--11, cited By 67.
\newblock \href {https://doi.org/10.1109/ICDCS.2009.76}
  {\path{doi:10.1109/ICDCS.2009.76}}.

\bibitem{Vazquez201468}
S.~Vazquez, T.~Muñoz-García, I.~Campanella, M.~Poch, B.~Fisas, N.~Bel,
  G.~Andreu, A classification of user-generated content into consumer decision
  journey stages, Neural Networks 58 (2014) 68--81, cited By 24.
\newblock \href {https://doi.org/10.1016/j.neunet.2014.05.026}
  {\path{doi:10.1016/j.neunet.2014.05.026}}.

\bibitem{Cheng201611}
G.~Cheng, J.~Han, A survey on object detection in optical remote sensing
  images, ISPRS Journal of Photogrammetry and Remote Sensing 117 (2016) 11--28,
  cited By 505.
\newblock \href {https://doi.org/10.1016/j.isprsjprs.2016.03.014}
  {\path{doi:10.1016/j.isprsjprs.2016.03.014}}.

\bibitem{Litjens201760}
G.~Litjens, T.~Kooi, B.~Bejnordi, A.~Setio, F.~Ciompi, M.~Ghafoorian,
  J.~van~der Laak, B.~van Ginneken, C.~Sánchez, A survey on deep learning in
  medical image analysis, Medical Image Analysis 42 (2017) 60--88, cited By
  3406.
\newblock \href {https://doi.org/10.1016/j.media.2017.07.005}
  {\path{doi:10.1016/j.media.2017.07.005}}.

\bibitem{Su2019828}
J.~Su, D.~Vargas, K.~Sakurai, One pixel attack for fooling deep neural
  networks, IEEE Transactions on Evolutionary Computation 23~(5) (2019)
  828--841, cited By 280.
\newblock \href {https://doi.org/10.1109/TEVC.2019.2890858}
  {\path{doi:10.1109/TEVC.2019.2890858}}.

\bibitem{Hornik1991251}
K.~Hornik, Approximation capabilities of multilayer feedforward networks,
  Neural Networks 4~(2) (1991) 251--257, cited By 2585.
\newblock \href {https://doi.org/10.1016/0893-6080(91)90009-T}
  {\path{doi:10.1016/0893-6080(91)90009-T}}.

\bibitem{raymond2020deep}
S.~J. Raymond, D.~J. Collins, R.~O’Rorke, M.~Tayebi, Y.~Ai, J.~Williams, A
  deep learning approach for designed diffraction-based acoustic patterning in
  microchannels, Scientific reports 10~(1) (2020) 1--12.

\bibitem{montgomery2020shale}
J.~Montgomery, S.~Raymond, F.~O’Sullivan, J.~Williams, Shale gas production
  forecasting is an ill-posed inverse problem and requires regularization,
  Upstream Oil and Gas Technology 5 (2020) 100022.

\bibitem{zhan2020deep}
X.~Zhan, Y.~Liu, S.~J. Raymond, H.~V. Alizadeh, A.~G. Domel, O.~Gevaert,
  M.~Zeineh, G.~Grant, D.~B. Camarillo, Deep learning head model for real-time
  estimation of entire brain deformation in concussion, arXiv preprint
  arXiv:2010.08527 (2020).

\bibitem{domel2021}
A.~G. Domel, S.~J. Raymond, C.~Giordano, Y.~Liu, S.~A. Yousefsani, M.~Fanton,
  N.~J. Cecchi, O.~Vovk, I.~Pirozzi, A.~Kight, B.~Avery, A.~Boumis, T.~Fetters,
  S.~Jandu, W.~M. Mehring, S.~Monga, N.~Mouchawar, I.~Rangel, E.~Rice, P.~Roy,
  S.~Sami, H.~Singh, L.~Wu, C.~Kuo, M.~Zeineh, G.~G. . D.~B. Camarillo, A new
  open-access platform for measuring and sharing mtbi data 11~(7501).
\newblock \href {https://doi.org/https://doi.org/10.1038/s41598-021-87085-2}
  {\path{doi:https://doi.org/10.1038/s41598-021-87085-2}}.

\bibitem{raymond2020towards}
S.~J. Raymond, J.~Maragh, A.~Masic, J.~R. Williams, Towards an understanding of
  the chemo-mechanical influences on kidney stone failure via the material
  point method, Plos one 15~(12) (2020) e0240133.

\bibitem{collins2020acoustic}
D.~Collins, S.~Raymond, Y.~Ai, J.~Willams, R.~O'Rorke, M.~Tayebi, Acoustic
  field design in microfluidic geometries via huygens-fresnel diffraction and
  deep neural networks, The Journal of the Acoustical Society of America
  148~(4) (2020) 2707--2707.

\bibitem{zhan2020prediction}
X.~Zhan, Y.~Li, Y.~Liu, A.~G. Domel, H.~V. Alidazeh, S.~J. Raymond, J.~Ruan,
  S.~Barbat, S.~Tiernan, O.~Gevaert, et~al., Prediction of brain strain across
  head impact subtypes using 18 brain injury criteria, arXiv preprint
  arXiv:2012.10006 (2020).

\bibitem{liu2021time}
Y.~Liu, A.~G. Domel, N.~J. Cecchi, E.~Rice, A.~A. Callan, S.~J. Raymond,
  Z.~Zhou, X.~Zhan, M.~Zeineh, G.~Grant, et~al., Time window of head impact
  kinematics measurement for calculation of brain strain and strain rate in
  american football, arXiv preprint arXiv:2102.05728 (2021).

\bibitem{Ganin2016}
Y.~Ganin, E.~Ustinova, H.~Ajakan, P.~Germain, H.~Larochelle, F.~Laviolette,
  M.~Marchand, V.~Lempitsky, Domain-adversarial training of neural networks,
  Journal of Machine Learning Research 17, cited By 1505 (2016).

\bibitem{raymond2014meshfree}
S.~Raymond, V.~Lemiale, R.~Ibrahim, R.~Lau, A meshfree study of the
  kalthoff--winkler experiment in 3d at room and low temperatures under dynamic
  loading using viscoplastic modelling, Engineering Analysis with Boundary
  Elements 42 (2014) 20--25.

\bibitem{raymond2015coupled}
S.~Raymond, Y.~Aimene, J.~Nairn, A.~Ouenes, et~al., Coupled fluid-solid
  geomechanical modeling of multiple hydraulic fractures interacting with
  natural fractures and the resulting proppant distribution, in: SPE/CSUR
  Unconventional Resources Conference, Society of Petroleum Engineers, 2015.

\bibitem{raymond2015estimation}
S.~Raymond, E.~Aimene, A.~Ouenes, et~al., Estimation of the propped volume
  through the geomechanical modeling of multiple hydraulic fractures
  interacting with natural fractures, in: SPE Asia Pacific Unconventional
  Resources Conference and Exhibition, Society of Petroleum Engineers, 2015.

\bibitem{raymond2016strategy}
S.~J. Raymond, B.~Jones, J.~R. Williams, A strategy to couple the material
  point method (mpm) and smoothed particle hydrodynamics (sph) computational
  techniques, Computational Particle Mechanics (2016) 1--10.

\bibitem{raymond2019modeling}
S.~J. Raymond, B.~D. Jones, J.~R. Williams, Modeling damage and plasticity in
  aggregates with the material point method (mpm), Computational Particle
  Mechanics 6~(3) (2019) 371--382.

\bibitem{wieghold2019detection}
S.~Wieghold, Z.~Liu, S.~J. Raymond, L.~T. Meyer, J.~R. Williams, T.~Buonassisi,
  E.~M. Sachs, Detection of sub-500-$\mu$m cracks in multicrystalline silicon
  wafer using edge-illuminated dark-field imaging to enable thin solar cell
  manufacturing, Solar Energy Materials and Solar Cells 196 (2019) 70--77.

\bibitem{raymond2020fracture}
S.~J. Raymond, B.~D. Jones, J.~R. Williams, Fracture shearing of
  polycrystalline material simulations using the material point method,
  Computational Particle Mechanics (2020) 1--14.

\bibitem{Harris2014623}
I.~Harris, P.~Jones, T.~Osborn, D.~Lister, Updated high-resolution grids of
  monthly climatic observations - the cru ts3.10 dataset, International Journal
  of Climatology 34~(3) (2014) 623--642, cited By 3789.
\newblock \href {https://doi.org/10.1002/joc.3711}
  {\path{doi:10.1002/joc.3711}}.

\bibitem{Meyers2006427}
M.~Meyers, A.~Mishra, D.~Benson, Mechanical properties of nanocrystalline
  materials, Progress in Materials Science 51~(4) (2006) 427--556, cited By
  3265.
\newblock \href {https://doi.org/10.1016/j.pmatsci.2005.08.003}
  {\path{doi:10.1016/j.pmatsci.2005.08.003}}.

\bibitem{Springer2006230}
R.~Springer, D.~Lowenthal, B.~Rountree, V.~Freeh, Minimizing execution time in
  mpi programs on an energy-constrained, power-scalable cluster, Vol. 2006,
  2006, pp. 230--238, cited By 66.
\newblock \href {https://doi.org/10.1145/1122971.1123006}
  {\path{doi:10.1145/1122971.1123006}}.

\bibitem{RAISSI2019686}
M.~Raissi, P.~Perdikaris, G.~Karniadakis,
  \href{https://www.sciencedirect.com/science/article/pii/S0021999118307125}{Physics-informed
  neural networks: A deep learning framework for solving forward and inverse
  problems involving nonlinear partial differential equations}, Journal of
  Computational Physics 378 (2019) 686--707.
\newblock \href {https://doi.org/https://doi.org/10.1016/j.jcp.2018.10.045}
  {\path{doi:https://doi.org/10.1016/j.jcp.2018.10.045}}.
\newline\urlprefix\url{https://www.sciencedirect.com/science/article/pii/S0021999118307125}

\bibitem{Zanna2020}
L.~Zanna, T.~Bolton, Data-driven equation discovery of ocean mesoscale
  closures, Geophysical Research Letters 47~(17) (2020) e2020GL088376,
  e2020GL088376 10.1029/2020GL088376.
\newblock \href
  {http://arxiv.org/abs/https://agupubs.onlinelibrary.wiley.com/doi/pdf/10.1029/2020GL088376}
  {\path{arXiv:https://agupubs.onlinelibrary.wiley.com/doi/pdf/10.1029/2020GL088376}},
  \href {https://doi.org/https://doi.org/10.1029/2020GL088376}
  {\path{doi:https://doi.org/10.1029/2020GL088376}}.

\bibitem{GOSWAMI2020102447}
S.~Goswami, C.~Anitescu, S.~Chakraborty, T.~Rabczuk,
  \href{https://www.sciencedirect.com/science/article/pii/S016784421930357X}{Transfer
  learning enhanced physics informed neural network for phase-field modeling of
  fracture}, Theoretical and Applied Fracture Mechanics 106 (2020) 102447.
\newblock \href {https://doi.org/https://doi.org/10.1016/j.tafmec.2019.102447}
  {\path{doi:https://doi.org/10.1016/j.tafmec.2019.102447}}.
\newline\urlprefix\url{https://www.sciencedirect.com/science/article/pii/S016784421930357X}

\bibitem{HAGHIGHAT2021113552}
E.~Haghighat, R.~Juanes,
  \href{https://www.sciencedirect.com/science/article/pii/S0045782520307374}{Sciann:
  A keras/tensorflow wrapper for scientific computations and physics-informed
  deep learning using artificial neural networks}, Computer Methods in Applied
  Mechanics and Engineering 373 (2021) 113552.
\newblock \href {https://doi.org/https://doi.org/10.1016/j.cma.2020.113552}
  {\path{doi:https://doi.org/10.1016/j.cma.2020.113552}}.
\newline\urlprefix\url{https://www.sciencedirect.com/science/article/pii/S0045782520307374}

\bibitem{HAGHIGHAT2021113741}
E.~Haghighat, M.~Raissi, A.~Moure, H.~Gomez, R.~Juanes,
  \href{https://www.sciencedirect.com/science/article/pii/S0045782521000773}{A
  physics-informed deep learning framework for inversion and surrogate modeling
  in solid mechanics}, Computer Methods in Applied Mechanics and Engineering
  379 (2021) 113741.
\newblock \href {https://doi.org/https://doi.org/10.1016/j.cma.2021.113741}
  {\path{doi:https://doi.org/10.1016/j.cma.2021.113741}}.
\newline\urlprefix\url{https://www.sciencedirect.com/science/article/pii/S0045782521000773}

\bibitem{HALL2021110192}
E.~J. Hall, S.~Taverniers, M.~A. Katsoulakis, D.~M. Tartakovsky,
  \href{https://www.sciencedirect.com/science/article/pii/S0021999121000875}{Ginns:
  Graph-informed neural networks for multiscale physics}, Journal of
  Computational Physics 433 (2021) 110192.
\newblock \href {https://doi.org/https://doi.org/10.1016/j.jcp.2021.110192}
  {\path{doi:https://doi.org/10.1016/j.jcp.2021.110192}}.
\newline\urlprefix\url{https://www.sciencedirect.com/science/article/pii/S0021999121000875}

\bibitem{HE2020103610}
Q.~He, D.~Barajas-Solano, G.~Tartakovsky, A.~M. Tartakovsky,
  \href{https://www.sciencedirect.com/science/article/pii/S0309170819311649}{Physics-informed
  neural networks for multiphysics data assimilation with application to
  subsurface transport}, Advances in Water Resources 141 (2020) 103610.
\newblock \href
  {https://doi.org/https://doi.org/10.1016/j.advwatres.2020.103610}
  {\path{doi:https://doi.org/10.1016/j.advwatres.2020.103610}}.
\newline\urlprefix\url{https://www.sciencedirect.com/science/article/pii/S0309170819311649}

\bibitem{JIN2021109951}
X.~Jin, S.~Cai, H.~Li, G.~E. Karniadakis,
  \href{https://www.sciencedirect.com/science/article/pii/S0021999120307257}{Nsfnets
  (navier-stokes flow nets): Physics-informed neural networks for the
  incompressible navier-stokes equations}, Journal of Computational Physics 426
  (2021) 109951.
\newblock \href {https://doi.org/https://doi.org/10.1016/j.jcp.2020.109951}
  {\path{doi:https://doi.org/10.1016/j.jcp.2020.109951}}.
\newline\urlprefix\url{https://www.sciencedirect.com/science/article/pii/S0021999120307257}

\bibitem{LIU2020113402}
M.~Liu, L.~Liang, W.~Sun,
  \href{https://www.sciencedirect.com/science/article/pii/S0045782520305879}{A
  generic physics-informed neural network-based constitutive model for soft
  biological tissues}, Computer Methods in Applied Mechanics and Engineering
  372 (2020) 113402.
\newblock \href {https://doi.org/https://doi.org/10.1016/j.cma.2020.113402}
  {\path{doi:https://doi.org/10.1016/j.cma.2020.113402}}.
\newline\urlprefix\url{https://www.sciencedirect.com/science/article/pii/S0045782520305879}

\bibitem{WANG2021109914}
S.~Wang, P.~Perdikaris,
  \href{https://www.sciencedirect.com/science/article/pii/S0021999120306884}{Deep
  learning of free boundary and stefan problems}, Journal of Computational
  Physics 428 (2021) 109914.
\newblock \href {https://doi.org/https://doi.org/10.1016/j.jcp.2020.109914}
  {\path{doi:https://doi.org/10.1016/j.jcp.2020.109914}}.
\newline\urlprefix\url{https://www.sciencedirect.com/science/article/pii/S0021999120306884}

\bibitem{YANG2021109913}
L.~Yang, X.~Meng, G.~E. Karniadakis,
  \href{https://www.sciencedirect.com/science/article/pii/S0021999120306872}{B-pinns:
  Bayesian physics-informed neural networks for forward and inverse pde
  problems with noisy data}, Journal of Computational Physics 425 (2021)
  109913.
\newblock \href {https://doi.org/https://doi.org/10.1016/j.jcp.2020.109913}
  {\path{doi:https://doi.org/10.1016/j.jcp.2020.109913}}.
\newline\urlprefix\url{https://www.sciencedirect.com/science/article/pii/S0021999120306872}

\bibitem{ZHANG2021100220}
Z.~Zhang, G.~X. Gu,
  \href{https://www.sciencedirect.com/science/article/pii/S2095034921000258}{Physics-informed
  deep learning for digital materials}, Theoretical and Applied Mechanics
  Letters (2021) 100220\href
  {https://doi.org/https://doi.org/10.1016/j.taml.2021.100220}
  {\path{doi:https://doi.org/10.1016/j.taml.2021.100220}}.
\newline\urlprefix\url{https://www.sciencedirect.com/science/article/pii/S2095034921000258}

\bibitem{ZOBEIRY2021104232}
N.~Zobeiry, K.~D. Humfeld,
  \href{https://www.sciencedirect.com/science/article/pii/S0952197621000798}{A
  physics-informed machine learning approach for solving heat transfer equation
  in advanced manufacturing and engineering applications}, Engineering
  Applications of Artificial Intelligence 101 (2021) 104232.
\newblock \href
  {https://doi.org/https://doi.org/10.1016/j.engappai.2021.104232}
  {\path{doi:https://doi.org/10.1016/j.engappai.2021.104232}}.
\newline\urlprefix\url{https://www.sciencedirect.com/science/article/pii/S0952197621000798}

\bibitem{doi:10.1122/8.0000138}
M.~Mahmoudabadbozchelou, M.~Caggioni, S.~Shahsavari, W.~H. Hartt,
  G.~Em~Karniadakis, S.~Jamali,
  \href{https://doi.org/10.1122/8.0000138}{Data-driven physics-informed
  constitutive metamodeling of complex fluids: A multifidelity neural network
  (mfnn) framework}, Journal of Rheology 65~(2) (2021) 179--198.
\newblock \href {http://arxiv.org/abs/https://doi.org/10.1122/8.0000138}
  {\path{arXiv:https://doi.org/10.1122/8.0000138}}, \href
  {https://doi.org/10.1122/8.0000138} {\path{doi:10.1122/8.0000138}}.
\newline\urlprefix\url{https://doi.org/10.1122/8.0000138}

\bibitem{Johnson2016694}
J.~Johnson, A.~Alahi, L.~Fei-Fei, Perceptual losses for real-time style
  transfer and super-resolution, Lecture Notes in Computer Science (including
  subseries Lecture Notes in Artificial Intelligence and Lecture Notes in
  Bioinformatics) 9906 LNCS (2016) 694--711, cited By 2211.
\newblock \href {https://doi.org/10.1007/978-3-319-46475-6-43}
  {\path{doi:10.1007/978-3-319-46475-6-43}}.

\bibitem{Larmor1883170}
J.~Larmor, On the direct application of the principle of least action to the
  dynamics of solid and fluid systems, and analogous elastic problems,
  Proceedings of the London Mathematical Society s1-15~(1) (1883) 170--185,
  cited By 5.
\newblock \href {https://doi.org/10.1112/plms/s1-15.1.170}
  {\path{doi:10.1112/plms/s1-15.1.170}}.

\bibitem{Agrawal2002368}
O.~Agrawal, Formulation of euler-lagrange equations for fractional variational
  problems, Journal of Mathematical Analysis and Applications 272~(1) (2002)
  368--379.
\newblock \href {https://doi.org/10.1016/S0022-247X(02)00180-4}
  {\path{doi:10.1016/S0022-247X(02)00180-4}}.

\end{thebibliography}

\end{document}